\documentclass[aps,pra,tightenlines,12pt,superscriptaddress,notitlepage,nofootinbib,longbibliography]{revtex4-1}

\usepackage{graphicx}  
\usepackage{xcolor}
\usepackage{dcolumn}   
\usepackage{bm}        
\usepackage{amssymb}   
\usepackage{amsmath}
\usepackage{verbatim}
\usepackage[hidelinks]{hyperref}
\usepackage[utf8]{inputenc}     
\usepackage{tabularx}
\usepackage{multirow}
\usepackage[capitalise,noabbrev]{cleveref}
\usepackage[centering,margin=1in]{geometry}
\usepackage{braket}
\usepackage{setspace}

\AtBeginDocument{%
    \newwrite\bibnotes
    \def\bibnotesext{Notes.bib}
    \immediate\openout\bibnotes=\jobname\bibnotesext
    \immediate\write\bibnotes{@CONTROL{REVTEX41Control}}
    \immediate\write\bibnotes{@CONTROL{%
    apsrev41Control,author="08",editor="1",pages="1",title="0",year="1"}}
     \if@filesw
     \immediate\write\@auxout{\string\citation{apsrev41Control}}%
    \fi
}%

\newcommand\snowmass{\begin{center}\rule[-0.2in]{\hsize}{0.01in}\\\rule{\hsize}{0.01in}\\
\vskip 0.1in Submitted to the  Proceedings of the US Community Study\\ 
on the Future of Particle Physics (Snowmass 2021)\\ 
\rule{\hsize}{0.01in}\\\rule[+0.2in]{\hsize}{0.01in} \end{center}}

\begin{document}

\title{Snowmass 2021 White Paper\\Computational needs of quantum mechanical calculations of materials for high-energy physics}

\author{Sin\'ead M.\ Griffin}
\affiliation{Materials Sciences Division, Lawrence Berkeley National Laboratory, Berkeley, California 94720, USA}
\affiliation{Molecular Foundry Division, Lawrence Berkeley National Laboratory, Berkeley, California 94720, USA}

\setstretch{1.2}

\begin{abstract}
Searches for new physics in high-energy physics experiments commonly rely on interactions with materials. A burgeoning direction is the accurate calculation and design of materials for high-energy physics (HEP) applications. In this Snowmass contribution, I briefly motivate the science need for quantum mechanical calculations of materials for HEP and outline the range of questions that such calculations can address. With this information, I assess the computational needs for \textit{ab initio} calculations in HEP, the specific computational resources and workflows used by state-of-the-art methods, and finally identify  promising future directions such as the use of machine learning and  strongly-correlated quantum mechanical calculations moving towards materials calculations on quantum computers. 

\end{abstract}

\snowmass

\maketitle

\newpage

\tableofcontents

\title{
Snowmas 2021 White Paper\\Computational needs of quantum mechanical calculations of materials for HEP applications
}

\section{Science Need for Advanced Calculations of Quantum Materials in HEP}

Next-generation HEP experiments are firmly entering the `quantum' realm where the energy scales and interactions of interest are approaching quantum limits\cite{Farhi_et_al:2015, BESHEP:2015, Chattopadhyay_et_al:2016}. This push towards quantum is aptly exemplified in searches for new physics (such as dark matter (DM)) that lie below the weak mass scale both in proposing new modes of coupling to the standard model, and fresh strategies for sensors and readout that surpass current sensing limits. Because of this, an in-depth understanding of the quantum behavior of the \textit{materials} that make up target/sensing systems is now needed for:
\begin{itemize}
  \item Accurate calculations of the material's form factor for specific couplings to new physics
  \item  Target/sensing efficiency calculations that incorporate decay channels, transduction, and carrier transport
  \item The role of materials inhomogeneities, defects and other decoherence channels for interpretation of near-threshold measurements
  \item Materials design for the systematic improvement of detector/sensor regimes
  \item Novel detection/sensing platforms based on (exotic) materials properties
\end{itemize}

As a brief motivational case study, we consider the need for materials theory and design through state-of-the-art quantum mechanical calculations in sub-GeV DM detection based on quasiparticle scattering (e.g. phonons)\cite{Essig:2016,Griffin:2019}. In such proposals, DM interacts with the standard model through quasiparticle scattering, with the type and profile of quasiparticle generated depending on the DM-matter coupling\cite{Trickle:2020}. With the known DM constraints on masses and interactions, the challenge lies in the accurate calculation of the material's response,  the evolution of these quasiparticles before and during readout, and their transduction efficiency for readout. Therefore, accurate quantum mechanical descriptions and design of materials' properties and function are required for any next-generation HEP detector or sensor relying on low-threshold phenomena\cite{Kahn/Lin:2022}. \\

First-principles descriptions of
such materials' properties, responses and energy transfer processes is possible using state-of-the-art Density Functional Theory (DFT) calculations\cite{Hafner_et_al:2006}. DFT is the work-horse for the \textit{ab initio} description of materials at a quantum mechanical level
and has recently been applied to accurately describe a range of HEP science problems such as DM-electron and DM-phonon interactions in a variety of materials\cite{Blanco_et_al:2021,Catena_et_al:2021,Griffin:2019, Marsh_et_al:2019, Roising_et_al:2021}. Crucially, being entirely first-principles these robust methods eliminate the need for empirical parameters, and when carefully applied, can describe and predict the behavior of materials within a few percent of accuracy. The power and accuracy of these approaches can also enable the inverse design of target functional properties, such as that recently carried out for the design of spin-orbit semiconductors for electron-based DM detection\cite{Inzani_et_al:2021}. \\

Applications of \textit{ab initio} approaches in HEP include the prediction of novel materials systems/phenomena for targets and sensors including new chemistries/materials, combinations of materials, and material architectures that can both maximize current detector/sensing efficiency and propose entirely new paradigms and systems with improved functionality. They can also be employed to accurately quantify material-based origins of excess signals, diagnose decoherence in quantum systems, and suggest mitigation strategies -- for example these methods have recently identified the origin of spin-based decoherence channels in superconducting qubits -- the resulting materials-based improvements led to a factor of 6 increase in resonator quality factor\cite{Altoe_et_al:2022}. Current applications of quantum materials in HEP focus on single-particle phenomena that are typically described in a non-interacting framework. However, these proposals do not take advantage of the thriving field of correlated quantum phenomena where small perturbations can result in threshold or cascade events, apt for small mass/energy detection\cite{Keimer/Moore:2017}. Looking to the future, contemporary discoveries in quantum materials of novel correlated phases and phenomena including spin, topology and orbital degrees of freedom should be explored for HEP applications, with the corresponding quantum mechanical calculations going beyond DFT to accurately include strongly-correlated physics (e.g. DFT+DMFT).

\section{Computational Efforts in Quantum Materials for HEP}

\subsection{Accurate description of quantum materials and phenomena}
Calculations of quantum materials including their single-particle spectra are currently used to estimate electronic and phononic properties of materials for HEP applications\cite{Griffin:2019}. Standard first-principles methods for this offer a balance of accuracy with computational efficiency, and can be extended to include higher-order diagrammatic expansions of quasiparticle interactions (e.g. multiphonon responses). This can also include the prediction  of quasiparticle transduction and coherence between disparate degrees of freedom (e.g. spin to charge) to propose novel detection and sensing schemes\cite{Roising_et_al:2021}. However, beyond-DFT approaches are required when phenomena beyond their mean-field, ground state properties are relevant, sometimes at a much higher computational cost. For example,  underlocalization in commonly-employed DFT functionals results in bandgaps and defect properties that can deviate significantly from experiment -- advanced methods such as the use of hybrid functionals, that include the calculation of exact exchange, can improve this with a higher compute cost. The accurate description of quasiparticle excitations requires calculation of the self-energy of the many-body electron problem through the GW approximation\cite{GW}, for example, resulting in a significant increase in computational cost. Strong correlations beyond those captured with simple extensions to DFT can be incorporated through the construction of embedding theories such as Dynamical Mean-Field Theory (DMFT)\cite{Paul/Birol:2019}, which again come at a high computational cost. Looking to the future, quantum computers offer great potential for simulations of such  quantum embedding models (e.g. the Hubbard model), with nearer-term applications using hybrid quantum-classical approaches to simulate materials' properties\cite{Ma_et_al:2020}.

\subsection{Materials informatics for accelerated discovery and inverse design}
Materials theory and design is enhanced by the construction of materials databases (the Materials Genome Initiative\cite{MGI:2018}) facilitated by both improved quantum mechanical methodology and high-performance computing (HPC) capabilities. Using materials informatics approaches, novel chemistries and materials can be identified from these large databases (e.g. the Materials Project\cite{MP}) by combining targeted high-throughput calculations, materials theory, data science, and machine learning, and have made key discoveries in fields as disparate as batteries, superconductivity and photovoltaics. Nascent efforts to use such computational materials informatics approaches are being explored for novel HEP detector materials \cite{Geilhufe_et_al:2018, Inzani_et_al:2021}, and have a huge potential for materials and systems optimization in HEP. Such identification and prediction of materials can then be used as a guide for the synthesis of optimal sensor and detector materials, with a theory/experiment feedback loop for improved predictions and synthesis. For example, following the proposal of Marsh et al. to detect axion DM in antiferromagnetic topological insulators (AFTIs)\cite{Marsh_et_al:2019}, several new AFTIs were identified by combining a high-throughput search of the Materials Project with DFT calculations and symmetry analysis\cite{Frey_et_al:2020}. Automation of high-throughput calculations and searches can accelerate discovery through workflows developed for materials theory applications, towards optimization of target and response properties. These results can be made available to the community through the publication of online databases for general reference datasets.

\subsection{Machine learning (ML) for quantum mechanical calculations}
Full \textit{ab initio} treatment of materials properties encompasses a range of established methods as described above, with the computational cost being  proportional to the level of complexity and accuracy needed (see Section~\ref{resource}). A rapidly emerging approach is the development of machine learning models to reproduce the essential features of these complex calculations but with significantly reduced computational overhead\cite{Burke:2020}. Training data for ML models will require the production and curation of high-quality calculations (10s-100s of GBs) which can then be centralized and disseminated for a variety of applications in quantum materials in addition to HEP.

\section{General Patterns of HPC Usage}
\subsection{Resource requirements}\label{resource}

\begin{figure}
  \includegraphics[width=0.3\textwidth]{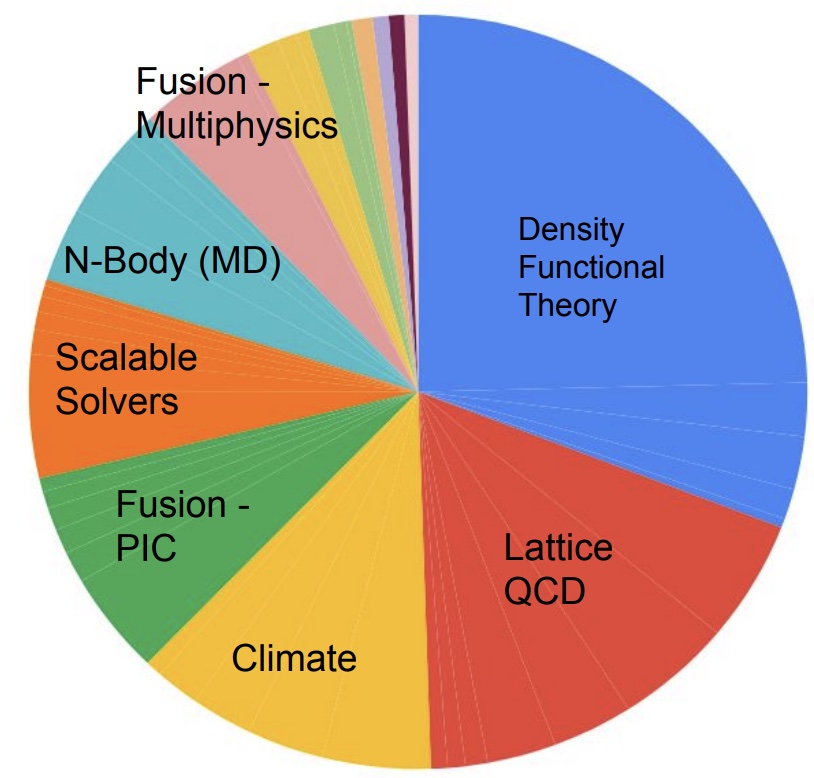}
  \caption{Top algorithms among NERSC codes for allocation year 2018. Reproduced from Ref.~\cite{NERSC:2018}}
  \label{fig:NERSC}
\end{figure}

The most used quantum mechanical first principles method is DFT which, for example, accounts for over 70\% of NERSC allocation time in materials science/solid state physics, and is the most used algorithm at NERSC (for comparison see DFT and Lattice QCD statistics for NERSC AY 2018 in Fig.\ref{fig:NERSC}).  A standard DFT calculation solves the many-body Schr\"{o}dinger equation using careful approximations -- most common DFT algorithms have O(N$^3$) scaling where N is the number of electrons in the system. Therefore, computational requirements can reach very large scales with increasing system sizes and/or complexity (e.g. accurate simulations of interfaces of multi-component systems). While a typical calculation solves the many-body Hamiltonian for total energy in a single self-consistent calculation, more complex calculations needing several self-consistent calculations and/or derivatives of energy to obtain forces (e.g. to find optimal atomic geometries) can rapidly increase the computational resources needed. Commonly, DFT practitioners are limited to a few hundred atoms for standard simulations because of the O(N$^3$) scaling -- many strategies for dealing with more complex larger systems are used including splitting up large models into smaller subsystems, or developing effective models derived from \textit{ab initio} parameters. However, for some cases, petascale (and even exascale) resources may be required, for example when studying complex heterostructures that cannot be simplified further, or for materials and phenomena that require higher-level theory and thus higher compute resources (e.g. the GW approximation has applications with O(N$^4$) scaling\cite{delBen:2019}). More recently, O(N) scaling algorithms -- linear scaling DFT -- have been developed and implemented in code packages\cite{linearDFT, ONETEP}, but at the expense of some of the range of applicability and accuracy for the systems under study, and are generally not currently suitable for correlated quantum materials that are the main focus of HEP applications. 

Workflows for calculations of materials make use of a wide variety of HPC resources, ranging from few-node local clusters and mid-sized institution compute clusters to larger scale national facilities such as NERSC and XSEDE. In addition to conventional CPU architectures, several standard code packages are now optimized and scale well with GPUs. Most practitioners do not use any of these resources exclusively with typical workflows requiring a combination for file preparation and testing (local), and larger production runs (national HPC facilities). While some analysis can be performed on-the-fly, most workflows will make extensive use of a combination of HPC resources for preparation, analysis and curation of datasets. Because of this, efficient transfer of large data sets between these different compute resources is needed both during production runs and following the completion of the project for storage and archiving (often datasets may not reside on the HPC facilities where they have been created for future access, curation and re-use). Globus file transfer, for example, is often used for transfer between national facilities and local university clusters.


\subsection{Data generation, transfer, curation and dissemination}
Workflows using DFT and beyond-DFT methods to calculate materials properties typically generated 10s-100s TB of data per user, depending on the complexity of the system and the targeted properties. Much of this will be high-quality materials data that can be used for a wide range of applications both within and outside of HEP. To avoid duplication of efforts, and to maximize the utility of these computationally demanding calculations, they should be disseminated and maintained in databases that can be easily accessed and interpreted across disciplines. For example, the storage and dissemination of  raw data can be done with the DOE's Materials Data Facility (MDF)\cite{MDF}, which will also attach a unique identifier to each dataset for property attribution and provenance. Higher-level raw data which often comprises of extracted physically-relevant information can be curated and preserved through  \textit{MPContribs}\cite{MPContribs} as part of the Materials Project. Codes, models, model parameters and algorithms used in the  simulations, theoretical models and ML approaches should be made open source through common repositories such as github, with source datasets linked at their location in the MDF.  ML products (code, models and networks) can be disseminated through the MLExchange\cite{MLExchange}, a DOE platform for storing, maintaining and tracking ML-related data. 

\section{Outlook}
Given the vast potential for the calculation, prediction and design of materials for HEP applications, this whitepaper outlines the computational needs for such efforts, and considerations for planning future HPC resources that take these into account. This effort also highlights the need to multidisciplinary collaboration across HEP, condensed matter physics and computational materials science, and how common datasets, codes and materials properties can be an effective means of accelerated progress in the design and interpretation of HEP experiments.

\bibliography{refs}

\end{document}